**Complexity Induced Sporadic Localized Multifractal Antiscreening in Gravitational Evolution at Large Scales**

Tom T.S. Chang, Kavli Institute for Astrophysics and Space Research, Massachusetts Institute of Technology, Cambridge, MA 02139, USA

**Abstract**

It has been suggested that antiscreening effects due to the running of the gravitational constant $G$ might provide a partial solution to the dark matter mystery. It has also been hypothesized that renormalization group scaling transformations at large scales might supply the theoretical explanation. In this letter, we demonstrate that multifractal coarse-graining scaling effects due to classical fluctuations in the IR with consecutive symmetry breakings in gravitational evolution and induced running of the gravitational constant with fractal structures at larger scales may provide the plausible explanation of the observed results of weak lensing observations and beyond. The sporadic and localized antiscreening due to the running of the gravitational constant can also provide the backbone for the cosmic evolution and large scale structure formation. Our interpretation of this interesting finding is that such effects are the result of the complexity phenomenon involving the evolution of large-scale multifractal structures and accompanying fluctuations, not the conventional arguments suggesting quantum gravity being the primary cause. We also suggest that the running of the cosmological constant due to such stochastic complexity evolution may provide a key to the understanding of the observed cosmic acceleration.

The concordance $\Lambda CDM$ model of cosmic evolution includes a cold dark matter component in an effort to explain away the observed anomalous rotational curves of the galaxies and gravitational lensing of light by clusters of galaxies, and to provide the gravitational backbone for the structure formation. It also includes a small positive cosmological constant to account for the accelerating expansion of the Universe [1,2]. It has been suggested [3-5 and references contained therein] that antiscreening effects due to the running of the gravitational constant might provide a partial alternative substitute for the dark matter assumption. It has also been hypothesized that renormalization group (RG) scaling transformations at large scales might supply the theoretical explanation.

In this letter, we demonstrate that RG multifractal scaling effects in the IR can yield such results. Our interpretation of this interesting finding is that these effects result mainly from the complexity phenomenon involving the coarse-graining of classical fluctuations, symmetry breakings, and evolution of large-scale multifractal structures. The same approach should also be capable of providing a plausible explanation of the observed acceleration of cosmic expansion due to the running of the cosmological constant.

There are a number of exact RG formulations in the literature [6-13]. Our starting point is the exact, non-perturbative, one-particle-irreducible RG formulation based on the scale-dependent effective action introduced by Nicoll and Chang [8] for static critical phenomena and Chang et al. [9] for dynamical systems far from equilibrium. This theory and its applications have been comprehensively reviewed in a Physics Report [11]. In the application of this effective field theory, we take the point of view that even though currently there are uncertainties of quantum gravitational UV closure, it does give us a reasonable argument that the gravitational and cosmological constants in the theory of general relativity are running (i.e., scale-evolving) coupling constants. During the matter dominated era, however, they evolve mainly due to the coarse-graining of classical fluctuations and the sporadic localized running of the coupling constants of the effective field as a function of the momentum "**k**" according to the Wilsonian RG equation as described below.

Since this paper is addressed to both the field theorists and condensed matter physicists, we shall use nomenclatures of both disciplines in the following description. Consider the partition function $Z$ of a fluctuating scalar field (order parameter) $\psi$ and define a generating functional for the n-particle correlation functions, $\overline{G}$ (Gibbs free energy) as

$$\exp[-\overline{G}(J_k)] = Z(J_k) = \prod_k \int d\psi_k \exp\left[-H(\psi_k) + \sum J_{-k}\psi_k\right] \quad (1)$$

where $H(\psi_k)$ is the classical action (Hamiltonian), $J_k$ is a source term (ordering field) and the expectation value $\overline{\psi}_k \equiv \langle\psi_k\rangle = -\delta\overline{G}/\delta J_{-k}$. The Legendre transform then gives the effective action $\overline{A}(\overline{\psi}_k)$ (Helmholtz free energy) which is a generating functional of the one-particle irreducible n-point Green functions:

$$\overline{A}(\overline{\psi}_k) = \overline{G} + \sum J_{-k}\overline{\psi}_k \quad (1)$$

We are interested in understanding the evolution of a complex dynamic system (the cosmic fluid) by viewing its coarse-grained prescription including classical fluctuations at larger and larger scales. To obtain a RG generator for such a procedure, we derive a differential functional equation for the scale-dependent effective action

$A(\psi_k, \psi_q, \overline{\psi}_p)$ defined in analogy to (1) by saddle point functional integrations of $\psi_k$ fluctuations within infinitesimal momentum (**k**)-space shells successively inward, where **q** represents the present momentum shell of integration and **p** the shells of prior integrations (with $\overline{\psi}'_p = \overline{\psi}_p(\psi_k, \overline{\psi}_q)$ evaluated at the $q$ shell saddle point). These shells are required to only collectively span the range of **k** although for simplicity in discussion below we employ the notation corresponding to isotropic shells. We suppose saddle point integrations have been performed over the range of $|\mathbf{k}| > 1$ (denoted as **p**-shells) and now functionally integrate over a thin shell $(1 - \Delta \ell) < |\mathbf{k}| \leq 1$ (denoted as **q**-shell) and evaluate appropriately the incremental change $\Delta A$ of the scale-dependent effective action while making sure that prior thin shell integrations preserve their saddle point characteristics. In the limit of $\Delta \ell \to 0$, we obtain the desired differential generator [8,9,11]. In the general case of $\psi$ (e.g., for vectors or tensors, time dependence, and/or with other fields that couple to $\psi$), we may merely in the formulation include additional indices such as $\alpha, \beta, \cdots$ (each of which results in a trace with $A$ now a matrix). We have, thus, the generalized generator as:

$$\frac{\partial A}{\partial \ell} = \frac{1}{2} \int \frac{d\Omega}{(2\pi)^d} \mathrm{tr} \ln \left( A_{q,-q} - \int \frac{dp}{(2\pi)^d} \int \frac{dp'}{(2\pi)^d} A_{qp} \left( A^{-1} \right)_{pp'} A_{p',-q} \right) \quad (2)$$

(with the additional indices suppressed for clarity) where $d\Omega$ is the differential element of the surface defined by $|\mathbf{q}| = 1$, $A_{qp} = \delta^2 A / \delta \psi_q \delta \overline{\psi}_p \big|_{\{\psi_q\} = \{\overline{\psi}_q\}}$ and analogous expressions for the other Hessians, $A^{-1}$ is the inverse of the matrix $A$, and the trace (tr) is understood to include integrations over continuous indices.

The differential RG functional generator (2) is exact, non-perturbative, one-particle irreversible and in closed form. And it has been proven [8] explicitly as the infinitesimal form of the usual one loop term.

We shall now apply the RG generator to cosmological gravitational evolution during the process of formation of structures while treating the gravitational constant as a running parameter. As suggested in the introductory remarks, we shall consider such processes as the result of complexity phenomenon induced by stochastic fluctuations, symmetry breakings, and spatiotemporal multifractal intermittency in the classical GR regime (i.e., at spatial scales much larger than the Planck scale). Thus, we do not need to worry about the usual UV closure difficulty as in quantum gravity and the nonperturbative RG generator works appropriately in the range of interest including both exact and perturbative calculations.

Complexity phenomenon can set in from an interacting dynamical system composed of many individual elements. If the interactions are nonlinear and long-ranged, quite often large scale structures are generated – each (well developed or partially formed) structure being composed of many individual elements behaving more or less coherently together. Such "coherent structures" can have varied sizes. And the ensuing behavior resulting from the diversified interactions of these structures can become very complicated [14]. Such is the situation in dynamical gravitational evolution [15]. As the masses in the cosmo evolve gravitationally in time under the initial influence of small fluctuations, we expect structures to begin to form due to linear and nonlinear gravitational instabilities and accompanying stochastic fluctuations. These entities will then interact and produce further enhanced fluctuations as well as new smaller and larger structures, a phenomenon vividly demonstrated over and over again via large scale ab initio numerical simulations based on, e.g., the $\Lambda CDM$ model. Our suggestion here is that this inherent complexity phenomenon can provide the cause for the running of the gravitational and cosmological constants due to coarse-graining effects because of the classical fluctuations and may be understood from RG arguments. These effects, being generally sporadic and localized will overwhelm any probable quantum gravitational effects, if any, particularly at large scales.

Before we proceed to describe such complexity induced gravitational evolutionary scenario involving symmetry breakings in terms of multifractals, let us first consider the lowest level simplistic nonlinear coarse-graining effect due to the cooperative effects of small fluctuations with minimal direct mode-couplings. We work with the classical Einstein-Hilbert action and employ the RG generator of Nicoll and Chang [8] including that of the cosmological constant. We note here that classical theory of criticality admit fluctuations beyond the standard mean field theory and effects of such fluctuations generally overwhelm those due to the quantum effects except at extremely low temperatures. If all coupling effects of the gravitational constant $G$ with the ordering fields are neglected at this level, then the anomalous dimension of $G$ related to such couplings are not present. Thus, we have, for $d=4$, $\partial G/\partial \ell = \partial G/\partial \ln k \sim G^2 + \cdots$. This simple result has been obtained by Rodriques et al. [4] and others by noting that $\partial G^{-1}/\partial \ell = \partial G^{-1}/\partial \ln k = 2\nu G_0^{-1}$ from e.g., the $\overline{MS}$-based RG. Integrating this expression yields the logarithmic scaling formula of Shapiro et al. [3] $G(k) = G_0/(1+\nu \ln(k^2/k_0^2))$ where the subscripts denote the initial state and $\nu$ an adjustable constant. By making a further assumption that the scale $k$ for the dynamics of disk galaxies is related exponentially with the effective Newtonian potential, they have found that this type of logarithmic scaling may explain a number (nine) of selected observed galaxy rotation curves quantitatively without the introduction of dark matters. Our point here is that such logarithmic scaling may be induced by low-level classical fluctuations for gravitational evolution without invoking the concept of quantum gravity, although such effects are trivial and probably cannot be relied on to provide a universal explanation for all observed rotational curves of the galaxies. We mention this result here mainly as a side note.

We are interested in scales much larger than that of individual galaxies and will now proceed to consider the coupling of the running of the gravitational constant with the other dynamically running scaling parameters in gravitational evolution. The cosmo evolves at vastly different scales. At smaller scales, multifractal structures related to the nonlinear evolution of dynamical coupling constants in RG can set in and these fluctuations are akin to those commonly observed in hydrodynamic turbulence [15]. Such phenomena do not need the influence or existence of nearby fixed points in RG phase space.

At larger scales, instabilities and self-organizations set in due to localized gravitational collapses and condensations. These events can exhibit bona fide symmetry-breaking phenomena with associated fixed points similar to those observed in condensed matter physics. Because of the slow varying nature of the gravitational constant, it will be possible to talk about fixed points of lines of criticality and symmetry breakings in the sense that there are fixed points in the RG phase space where $G$ is taken as a slowly varying and irrelevant parameter. *"G"* is an irrelevant parameter along a line of criticality until it develops a cusp singularity with a particular fractal characteristic due to an induced linear or nonlinear instability (accompanied by a new fixed point) because of the mutual interaction of the running of the gravitational constant with the other RG coupling constants. At this point higher order crossover phenomenon sets in due to the competition of orderings of different Lifshitz characteristics [16]. Such phenomena occur quite frequently in condensed matter physics.

We are mainly interested in the gravitational evolution during the matter-dominated era within the domain of non-relativistic dynamic motion. Thus, we shall start with the Einstein-Hilbert effective action and work under the Newtonian approximation by assuming (1) the gravitational field is weak, (2) the variations of fields are slow in time, and (3) the particles are nonrelativistic. Therefore, the only surviving metric element is $g^{00} \approx 1 - h_{00}$ with $|h_{00}| \ll 1$. The approximate effective action is then [17]:

$$A = (8\pi G)^{-1} \int (\nabla \varphi)^2 d^3 x dt - \sum m \int \varphi dt + \sum (m/2) \int v^2 dt \qquad (3)$$

where the gravitational potential $\varphi = c^2 h_{00}/2$. For the moment, we have not included the cosmological constant but will briefly return to it later. To (3), we shall append other allowable higher order interaction terms and the corresponding coupling constants due to fluctuations through coarse-graining, stochasticity, and the RG transformation.

Consider the situation of symmetry breakings due to the appearance of fixed points in the ordering of the gravitational potential $\varphi$ under the influence of the running of the gravitational constant. This is the most important ordering and symmetry-breaking effect in gravitational evolution as can be seen from (3). Treating $G$ as an irrelevant parameter, it is known that there exist relevant symmetry breaking fixed points for the ordering of $\varphi$ (some perturbative and some exact). Even with $G$ as an irrelevant parameter, the running of its **k**-dependence nevertheless has a direct influence on the $\varphi$-ordering due to the fluctuations as it has a direct effect on the characteristics of the propagator. Thus, the critical propagator may take on the general form of $\sum_{i=1}^{J}|\mathbf{k}_i|^{\sigma_i}$ with generalized Lifshitz character, where each $\mathbf{k}_i$ is a $d_i$-dimensional vector and $\sigma_i$ are the corresponding propagator exponents. We can use (2) modified for anisotropic momentum shells and with appropriate differential scale changes to calculate the scaling exponents [11]. To demonstrate the richness of the existence of various fixed points, we give below the explicit expression [16] for the eigenvalues to first order in $\varepsilon_O \equiv \lambda_O$ for $d$ less than the borderline dimension $d_b$ determined by $\sum_{i=1}^{J} d_i/\sigma_i = O/(O-1)$, where $O$ is the order of phase transition.

$$\lambda_p' = \lambda_p - 2\varepsilon_O \langle O, p; p \rangle / \langle O, O, O \rangle \qquad (4)$$

with $\lambda_p \equiv \left[\sum_{i=1}^{J} d_i \sigma_> / \sigma_i \right](1-p) + p\sigma_>$ being the Gaussian eigenvalues (with $p$ corresponding to the $\varphi^{2p}$ term of the interaction polynomial expansion of the $\varphi$-coupling) and the inner products $\langle O, p; p \rangle$ of the Gaussian eigen-operators are expressible in terms of the binomial coefficients in closed form:

$$\langle O, p; p \rangle = \sum_{j=0}^{[O/2]} \binom{p}{j}\binom{p-1/2}{j}\binom{2p-2j}{O-2j} \qquad (5)$$

These fixed points are located at $H_O^* \sim \varepsilon_O Q_O$ where $Q_O$ is the corresponding Gaussian operator. These contain the well-known result of the Wilson-Fisher fixed point and all generalized Lifshitz character fixed points with various helical orders.

From (4), all scaling exponents of the coupling constants may be calculated. As the competition of ordering for $\varphi$ intensifies with the variation of the **k**-dependence of $G$, eventually $G$ may develop special cusp characteristics (due to linear or nonlinear interaction effects and instabilities) at nontrivial values of $G^*$'s with enhanced fractal characteristics with respect to **k**. At these cusp singularity-induced fixed points (where, e.g., $G - G^* \sim \exp(\lambda_G \ell) \sim |k - k^*|^{\lambda_G}$ for the isotropic essentially diagonalized case) higher order symmetry breakings may set in. Some of these events are quite well known in condensed matter physics, such as the tricritical points (TCP), bicritical/tetracritical points with or without helical ordering [11,16,18,19], and many of the relevant exponents are actually already contained in the fixed point eigenvalues that generated (4). Such symmetry breakings of higher ordering may sometimes set in at the borderline dimensions $d_b$, and the eigenvalues are determined by the appropriate Gaussian eigenvalues $\lambda_p$ with corrections that sometimes are fractional powers of the logarithms [11,19]. Cusp-like fractal variations of $G$ in **k** near $G^*$'s will induce the anomalous increasing ($G - G^*$ increases with the decrease $|k - k^*|$) and decreasing trends of the antiscreening effect that can be interpreted as virtual particles or dark matters and also as anomalies of the results of observations via weak-gravitational lensing, etc.

In addition to the ordering effects of $G$ related to $\varphi$ alone, there can be competing orderings with the other field variables such as $m$ and $v$ and accompanying higher order interaction terms. These symmetry-breaking effects can introduce further sporadic and localized (both fading-in and fading-out) antiscreening effects on $G$.

Gravitational evolution is spatiotemporal. Thus, we must also consider stochastic effects (linear and nonlinear critical dynamics) for dynamical systems far from equilibrium. We may employ the RG generator of Chang et al. [9] to consider such effects. Generally, when the stochastic effects due to the time-varying noises on the ordering and symmetry breakings need to be considered for a non-equilibrium dynamic system, the conjugate fields of the ordering fields due to the imposed noise must be generated and included in the scale-dependent stochastic effective action. Space prevents us from delving into the complicated details of such calculations. Instead, we consider the results of a simple example below to demonstrate the typical spatiotemporal effects in gravitational evolution.

Consider, for example, the case of $\varphi$ Lifshitz ordering with weak mode coupling and its dynamics driven by a stochastic noise characterized by an isotropic propagator $(k^\sigma + r + i\omega/Q_k)$ where $\sigma = 2L$ with $L$ being the Lifshitz character, $\omega$ the Fourier component of the time variable, and $Q_k \propto k^y$ the characteristics of the noise. We find nontrivial fixed points for $d < 2\sigma$ and $y \neq 0$ to first order in $\varepsilon \equiv 2\sigma - d$ with eigenvalues the same as those of the isotropic version of (4) and an additional dynamic exponent $z = \sigma + y - \eta$ (which measures the anisotropic scaling between time and space) where the correlation function exponent $\eta$ for $\sigma = 2L$ when $L$ is an integer is:

$$\eta = \varepsilon^2 [(-1)^{L+1} \Gamma^2(2L) / 27 L \Gamma(L) \Gamma(3L)] + O(\varepsilon^3) \tag{6}$$

and $z = \sigma + y$ when $L$ is not an integer.

When $y = 0$, on the other hand, the dynamic exponent is $z = \sigma - \eta + I$ with

$$I = \varepsilon^2 2^\sigma \Gamma(\sigma)/27 \int_0^\infty dt \left[ \int_0^\infty dx \exp(-tx^\sigma) J_{\sigma-1}(x) \right]^3 + O(\varepsilon^3) \tag{7}$$

We note from this example that classical spatiotemporal fluctuations, orderings, and symmetry breakings are generally anisotropic between space and time in gravitational evolution.

In the above fixed point analyses, we found that critical RG trajectories may accompany cusp-like singularities of the gravitational constant in momentum space. Actual RG evolutions of the cosmic medium need not follow such trajectories exactly. In fact, they generally are continuous paths that may evolve close to the critical trajectories. The antiscreening effects then vary up and down (increase and decrease) following these trajectories. If one follows such trajectories and integrate the RG functional integrodifferential equation (2), we obtain the full effective action (Helmholtz free energy) $A(\overline{\psi})$ of the local evolving cosmic medium [20]. Thus, nonlinear equation(s) of state which may be obtained from the integrated effective action (Helmholtz free energy) of the evolving cosmic medium cannot be assumed a priori because they change sporadically and locally from criticality to criticality and within the crossover domains depending on its complexity structure. Since the RG generator is nonlinear and singular, generally the integrated effective actions or Helmholtz free energies are singular functions of the global nonlinear invariants (in the form of nonlinear functions of the coupling constants). As an example, we give below the calculated RG zero loop free energy for the case of a tricritical point (TCP) for a $k^2$ - propagator for a scalar field $\varphi$ in exponentiated form in $\Phi \equiv \langle \varphi \rangle$ [11,19]:

$$A(\Phi) = \frac{1}{2}\left[\mu_1 - \frac{1}{2}\frac{\mu_2^2}{\mu_3}(Y^{-1/5} - 1)\right]\Phi^2 + \frac{\mu_2}{4!}Y^{2/5}\Phi^4 + \frac{\mu_3}{6!}Y\Phi^6 \qquad (8)$$

where $\mu_i = a_i(0)$ ($i = 1, 2, 3$) are the initial values of the first three Gaussian eigenfunction expansion coefficients $a_i(\ell)$ of the RG generator, and $Y$, the so-called "quadrature function" which characterizes the evolution along the RG trajectory is

$$Y^{-1} = 1 - (\mu_3/\lambda)(3,3,3)\ln\Gamma \qquad (9)$$

with $a_1(\ell) = \Gamma e^{\lambda\ell}$, and the triple inner product of the Gaussian operators given below:

$$(i, j, k) = \frac{i!\,j!\,2k!}{2i!\,2j!\,k!}\sum_m \binom{j - \frac{1}{2}}{m}\binom{k}{j-m}\binom{2j-2m}{3-2m} \qquad (10)$$

(N.B., the singular nature of (9) involving the fractional powers of the logarithm, $\ln\Gamma$.)

Detailed examples of such functions of multiple singularities for multifractal fixed points with general Lifshitz characters may be found in Chang et al. [11] and Nicoll and Chang [19]. Thus, in general, for the complex cosmic domain, it is not possible to assume a priori an equation of state. The nonlinear equation of state or free energy must be determined self-consistently with the stochastic dynamics locally.

An alternative point of view of the complex cosmic medium is that at each local state of the full RG space, the medium may be considered to be characterized by a local invariant (of the coupling constants including $G$) with a fractal singularity which varies from one locality to another according to the change of the RG parameters and therefore the local state. This is the point of view taken by the recently developed concept of rank-ordered multifractal analysis (ROMA) [14,15]. This technique introduced by Chang and Wu [21] determines explicitly the complexity of the medium in terms of an implicit spectrum expressing the local fractal dimension "s" as a nonlinear function of the local invariant "$Y$". The method is particularly useful for analyzing complex states such as the organization of the grand structure of the Universe and has been applied successfully to the analysis of the cosmic Baryonic medium obtained by large scale $\Lambda CDM$ simulations. [15]   Such type of analyses contains much more statistical information for comparison than those provided by the conventional spectrum densities and correlation function techniques.

In summary, we have demonstrated analytically that fixed points are abundant in the RG phase space that can provide the onset of symmetry breakings and crossovers of competing orderings with fractal variations of the gravitational constant during gravitational evolution under general spatiotemporal fluctuations and self-organization. The analysis is based on the scale-dependent effective action for the Newtonian approximation and classical complexity phenomena with the running of the gravitational constant without invoking the quantum gravitational effects.  The full effect of such complex phase transition-like behavior indicate that phenomena of sporadic and localized antiscreening effects and their collaborative behavior with the orderings and competitions of orderings of self-organized coherent structures exist under the general gravitational evolution with antiscreening produced virtual particles (or dark matter) and the development of large structures and multifractal fluctuations.  Such complex and multifractal behavior can be analyzed and compared based on refined statistical descriptions such as the recently developed rank-ordered multifractal analyses (ROMA) [14,21] based on both the results of realistic large scale numerical simulations with a running gravitational constant and actual observations such as those obtained from weak gravitational lensing and anisotropic CMB measurements.

Finally, in this letter, for clarify and simplicity of discussion, we have ignored the discussion of the running of the cosmological constant. We assumed that while the running of the Newtonian constant and symmetry breakings may have an influence on the running of the cosmological constant and thereby producing temporal variations (accelerations or decelerations) of the speed of cosmic expansion, the feedback of the running of the cosmological constant to that of the Newtonian constant is negligible. We shall consider the running of the cosmological constant due to its coupling with the evolution of the complex cosmic medium in a separate document.

This research is partially supported by a grant from NSF.